\documentclass[%
 reprint,
 superscriptaddress,
 amsmath,amssymb,
 aps,
 prstper,
]{revtex4-2}

\usepackage{graphicx}
\usepackage{dcolumn}
\usepackage{bm}
\usepackage{url}

\usepackage[breaklinks]{hyperref}

\begin{document}

\title{Regulatory Changes in Power Systems\\Explored with Explainable Artificial Intelligence}

\author{Sebastian Pütz}
\affiliation{
  Karlsruhe Institute of Technology, Institute for Automation and Applied Informatics (IAI), 76344 Eggenstein-Leopoldshafen, Germany
}
\affiliation{
  Institute for Theoretical Physics, University of Cologne, 50937 Köln, Germany
}
\affiliation{
  Forschungszentrum Jülich,
  Institute of Energy and Climate Research,
  Systems Analysis and Technology Evaluation (IEK-STE),
  52428 Jülich, Germany
}

\author{Johannes Kruse}
\affiliation{
  Forschungszentrum Jülich,
  Institute of Energy and Climate Research,
  Systems Analysis and Technology Evaluation (IEK-STE),
  52428 Jülich, Germany
}
\affiliation{
  Forschungszentrum Jülich, Institute of Energy and Climate Research, Energy Systems Engineering (IEK-10), 52428 Jülich, Germany
}
\affiliation{
  Institute for Theoretical Physics, University of Cologne, 50937 Köln, Germany
}

\author{Dirk Witthaut}
\affiliation{
  Forschungszentrum Jülich,
  Institute of Energy and Climate Research,
  Systems Analysis and Technology Evaluation (IEK-STE),
  52428 Jülich, Germany
}
\affiliation{
  Forschungszentrum Jülich, Institute of Energy and Climate Research, Energy Systems Engineering (IEK-10), 52428 Jülich, Germany
}
\affiliation{
  Institute for Theoretical Physics, University of Cologne, 50937 Köln, Germany
}

\author{Veit Hagenmeyer}
\affiliation{
  Karlsruhe Institute of Technology, Institute for Automation and Applied Informatics (IAI), 76344 Eggenstein-Leopoldshafen, Germany
}

\author{Benjamin Schäfer}
\affiliation{
  Karlsruhe Institute of Technology, Institute for Automation and Applied Informatics (IAI), 76344 Eggenstein-Leopoldshafen, Germany
}

\begin{abstract}
A stable supply of electrical energy is essential for the functioning of our society. Therefore, the electrical power grid's operation and energy and balancing markets are subject to strict regulations. As the external technical, economic, or social influences on the power grid change, these regulations must also be constantly adapted. 
However, whether these regulatory changes lead to the intended results is not easy to assess. 
Could eXplainable Artificial Intelligence (XAI) models distinguish regulatory settings and support the understanding of the effects of these changes?
In this article, we explore two examples of regulatory changes in the German energy markets for bulk electricity and for reserve power.
We explore the splitting of the German-Austrian bidding zone and changes in the pricing schemes of the German balancing energy market. 
We find that boosted tree models and feedforward neural networks before and after a regulatory change differ in their respective parametrizations. 
Using Shapley additive explanations, we reveal model differences, e.g. in terms of feature importances, and identify key features of these distinct models.
With this study, we demonstrate how XAI can be applied to investigate system changes in power systems. 
\end{abstract}

\maketitle

\section{Introduction}

Modern society is highly dependent on the reliable operation of the electrical power grid ~\cite{petermannWhatHappensBlackout2011}.
Thus, the energy system is highly regulated to ensure a secure electricity supply.
These regulations are constantly reviewed and, if necessary, adapted in order to cope with ever-changing external technical economic, or social drivers~\cite{kruseRevealingDriversRisks2021, kruseSecondaryControlActivation2022}. For example, the regulation of the German electricity systems, in particular the EEG (German Renewable Energy Sources Act), has been revised repeatedly in the last decade~\cite{federalministryforeconomicaffairsandclimateactionErneuerbareEnergienGesetzEEG2023}.
\begin{figure*}[t]
  \centering
  \includegraphics[width=0.91\linewidth]{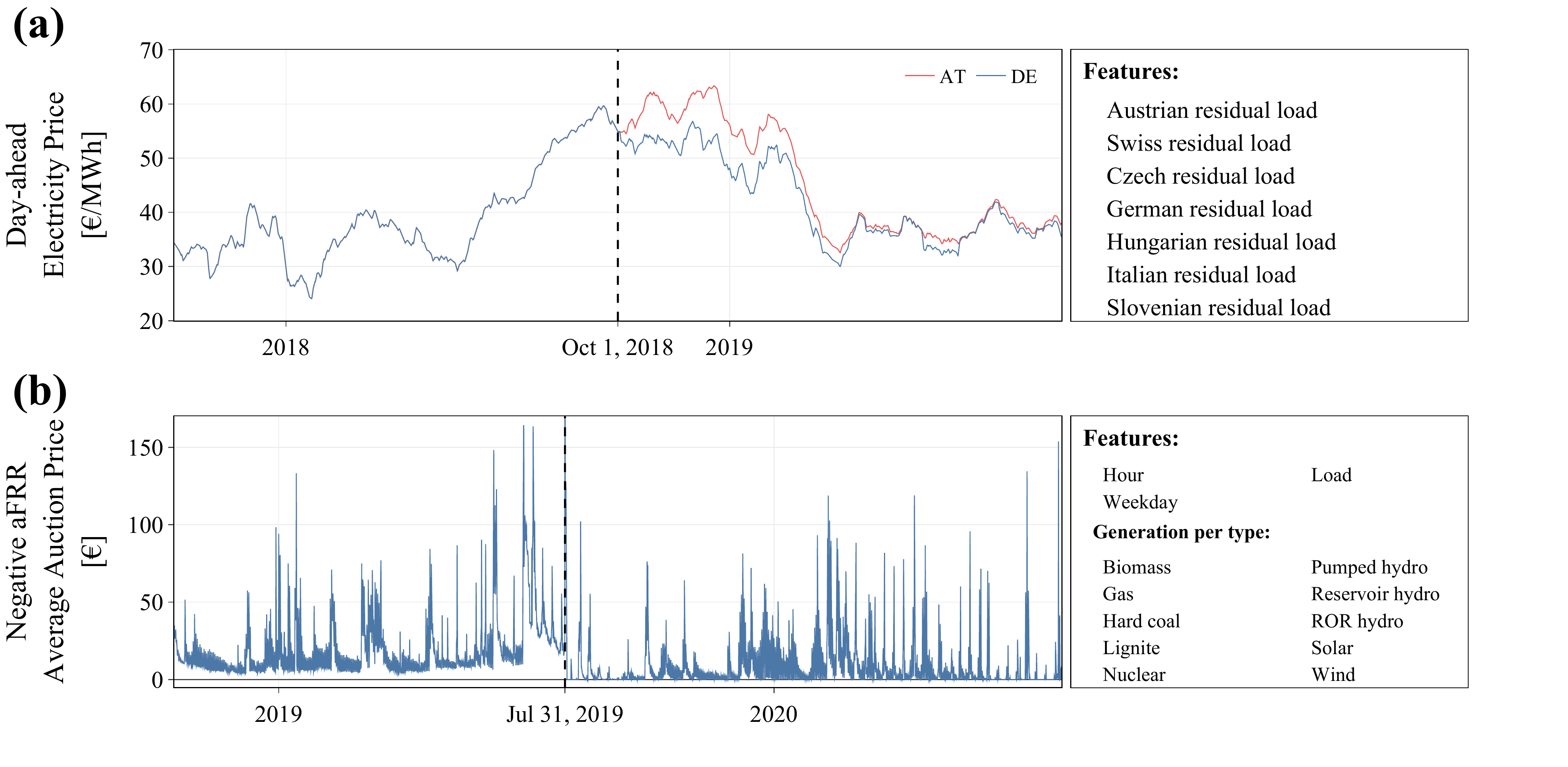}
  \caption{
    Regulatory changes in price time series.
    We develop regression models for price time series (left) using various external features (right) as inputs.
    (a) Time series of the day-ahead electricity price in Germany and Austria (depicted as a 30-day moving average for visibility). Before the bidding zone split (dashed vertical line), the two prices coincided. 
    (b) Time series of the average auction price for negative aFRR. Prices tended to be higher during the mixed price system (left of the dashed vertical line) than during the capacity price based auction.
    }
  \label{pricets}
\end{figure*}
Such regulatory changes require a careful ex-post evaluation to determine whether they serve their purpose or may have undesirable (side) effects. For instance, studies suggest that the past design of the German electricity markets incentivized market participants to be systematically short on energy, which had an undesirable effect on system stability \cite{ehrhartDesignRegulationBalancing2021, eickeElectricityBalancingMarket2021}. 
In the past decade, such effects of market design and regulatory changes have been increasingly studied using empirical methods \cite{kochShorttermElectricityTrading2019, weissbachImpactCurrentMarket2018, ockerGermanParadoxBalancing2017, ockerDesignEuropeanBalancing2016}. 
However, such an analysis is challenging as markets are affected by numerous agents, but publicly available data is often aggregated or anonymous~\cite{entso-eENTSOETransparencyPlatform, 50hertztransmissiongmbhRegelleistungNet}.

Modern Machine Learning (ML) tools provide powerful tools for the prediction of market dynamics from publicly available data. ML approaches were used to forecast electricity prices~\cite{lagoForecastingDayaheadElectricity2021, yang2022novel} and balancing market prices ~\cite{mertenAutomaticFrequencyRestoration2020}. Notably, modern ML models learn non-linear effects and interactions for high-performance predictions \cite{hastieElementsStatisticalLearning2016}, but they are often black-boxes, and therefore not directly applicable for the ex-post analysis of regulatory changes. Econometric analyses therefore mostly settle for linear models \cite{kochShorttermElectricityTrading2019, eickeElectricityBalancingMarket2021}. 

Here, we use tools from XAI to leverage modern ML methods for the evaluation of regulatory changes. XAI tools explain black-box models and therefore give insights into what the model has learned \cite{barredoarrietaExplainableArtificialIntelligence2020}. XAI in energy systems is a quickly growing field \cite{machlevExplainableArtificialIntelligence2022}, with applications ranging from power grid stability \cite{kruseRevealingDriversRisks2021} to price analysis \cite{trebbien2022understanding, moon2022toward}. Here, we explore the capabilities of XAI to distinguish regulatory settings and reveal effects of regulatory change on power system operation. Notably, this is fundamentally different from ML applications that \textit{detect} system drifts \cite{muschalikAgnosticExplanationModel2022, zimmermannImprovingDriftDetection2022} or anomalies \cite{turowski2022modeling, turowski2022enhancing}, as we investigate known changes. 

As a case study, we focus on two changes. First, we look at the split of the German-Austrian bidding zone for electricity prices. Second, we consider a change in the German balancing power market design. We model the market prices with Gradient-Boosted Trees (GBT) or Feedforward Neural Networks (FNN), as two distinct and prominent ML methods. To explore model changes, we explain the black-box models with the popular SHapley Additive exPlanation (SHAP) values~\cite{lundbergLocalExplanationsGlobal2020}.

This article is structured as follows: We first introduce the investigated regulatory changes (sec.~\ref{sec:intro_changes}), continue with a brief description of the applied XAI methods (sec.~\ref{methods}), and then present our results (sec.~\ref{results}) before closing with a conclusion and outlook (sec.~\ref{conclusion}).

\section{Two instances of regulatory changes}
\label{sec:intro_changes}
To study regulatory changes with XAI, we consider two recent market changes: A bidding zone split between Germany and Austria and a reform of the German balancing energy market. 
\subsection{Bidding Zone Split}
\label{bzs}
The European electricity market is organized in bidding zones. 
Bidding zones exhibit a uniform electricity price in their entire area (marginal pricing). 
Energy can be exchanged without capacity restrictions, i.e., it does not matter to a consumer whether the supplier is nearby or remotely located in the zone~\cite{RegulationEU20192019}.
Most bidding zones coincide with country borders. Some are smaller zones within a country and a few are even larger than a single country, for example, Germany, Luxembourg, and Austria shared a common bidding zone until October 2018. 

In this bidding zone, wind power resources are predominantly located in Northern Germany, while many industrial consumers are located in the Western and Southern parts of the bidding zone. The transport of electricity in the North-South direction repeatedly led to congestion of lines, which needs to be mitigated via TSO intervention.

To mitigate this congestion problem, the common bidding zone has been split~\cite{eexagVorbereitungTrennungStromgebotszone2018}.
As of October 1, 2018, Austria has been separated as a distinct bidding zone from Germany and Luxembourg.
As a result, consumers from Austria can no longer buy electricity from Germany without capacity limitations and the prices no longer coincide (see Fig~\ref{pricets}a).

\begin{figure*}[t]
  \centering
 \includegraphics[width=0.91\linewidth]{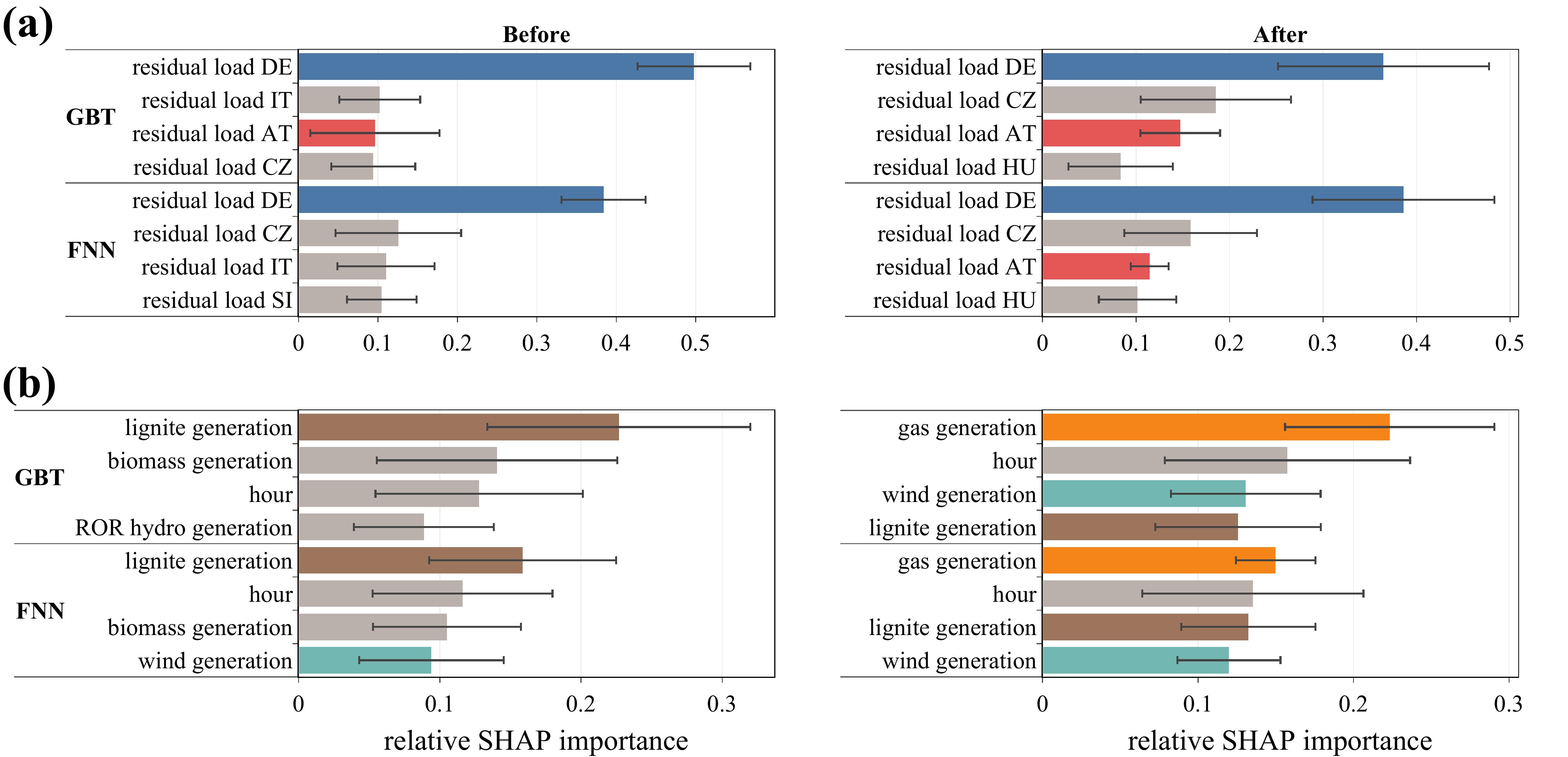}
  \caption{
  The models differ before and after a regulatory change. We visualize these changes in the models by plotting the relative SHAP importance of the most important features averaged over six models. The standard deviation of the relative SHAP importances is shown by the error bars.
  (a) Austrian residual load increases in importance relative to the german residual load after the bidding zone split. (b) Lignite decreases in importance while gas increases after the end of the mixed price system.
  }
  \label{importance_plot}
\end{figure*}

\subsection{Balancing Power Market Design}
\label{mpv}
Balancing power is essential for ensuring a stable electricity supply.
Generation and demand must always be in balance for the operation of a stable electricity grid~\cite{machowskiPowerSystemDynamics2020}.
Positive and negative balancing energy is therefore held in reserve to compensate for short-term imbalances. In the Continental European electricity grid, a three-stage system is in place for this purpose. 

The Frequency Control Reserve (FCR) is the primary control reserve that reacts within seconds to imbalances. For prolonged imbalances, the automatic Frequency Restoration Reserve (aFRR) as secondary control takes over in minutes and the manual Frequency Restoration Reserve (mFRR) as tertiary control takes over in a timescale of half-hours.
The amount of aFRR and mFRR required is assessed and tendered individually by each TSO. This is in contrast to FCR, where the tendered demand is set at the European level. Positive and negative aFRR and mFRR power are tendered separately~\cite{CommissionRegulationEU2017a}.
In the following, we are focusing on the german market design for aFRR.
Beginning in 2018, the daily aFRR capacity was tendered in six four-hour blocks.

For a single block, each market participant specifies what capacity they are willing to reserve for frequency control, together with a capacity price and an energy price. 
The capacity price is 
for the mere provision of the capacity. These costs are borne by the end consumers. The energy price compensates the supplier for the energy actually delivered upon activation. These costs are charged to the parties that are responsible for the imbalance.
a standby payment
Before October 2018, the bids were awarded solely based on the capacity price and independently of the energy price. The highest bid that is awarded sets the marginal capacity price.

From October 2018, the mixed price, i.e. a price composed of both the energy and the capacity price, replaced the capacity price in the supplier selection process~\cite{bundesnetzagenturBK618019BeschlussVom2018}.
The mixed price scheme was highly controversial and abolished again in July 2019~\cite{oberlandesgerichtdusseldorfKart806182019}(cf.~App.~\ref{app:mps}).
During all periods, the activation of aFRR reserves in case of a power imbalance is based on the energy prices alone~\cite{bundesnetzagenturFestlegungAusschreibungsbedingungenUnd2017}.
Both the energy price and the capacity price are settled in a pay-as-bid process.

\section{Methods}
\label{methods}
\subsection{Datasets}
To model the Austrian day-ahead electricity price before and after the bidding zone split we utilized the residual loads of all countries with which Austria has a cross-border connection.
We obtained day-ahead electricity prices and day-ahead forecasts for load and renewable generation via the ENTSO-E Transparency Platform~\cite{entso-eENTSOETransparencyPlatform}. Additionally, we retrieved the actual run-of-river (ROR) hydro generation. We created residual load forecast time series by subtracting the day-ahead wind and solar forecasts as well as a lagging run-of-river hydro generation average from the day-ahead load forecast. We utilized data for one year before and one year after the bidding zone split with an hourly resolution.

To explore the shift from mixed price auctions to capacity price-based auctions, we investigated negative aFRR
as a key control aspect and leave mFRR and positive aFRR for future work.
We focused on the average prices that were responsible for the acceptance of the bid in the respective auction schemes (see Fig~\ref{pricets}b). That is, we used the average mixed price from 16 October 2018 to 31 July 2019 and then the average capacity price for the period of one year. The reserve market data has been obtained from regelleistung.net~\cite{50hertztransmissiongmbhRegelleistungNet}. Due to the tendered block sizes, this data has a 4-hour resolution.
To model these prices, we have used the actual production by generation type in Germany as input features, which is also available at the ENTSO-E Transparency Platform~\cite{entso-eENTSOETransparencyPlatform}. Notably, data on actual bids is only available in an anonymized form~\cite{50hertztransmissiongmbhRegelleistungNet}.

\begin{figure*}[t]
  \centering
  \includegraphics[width=0.91\linewidth]{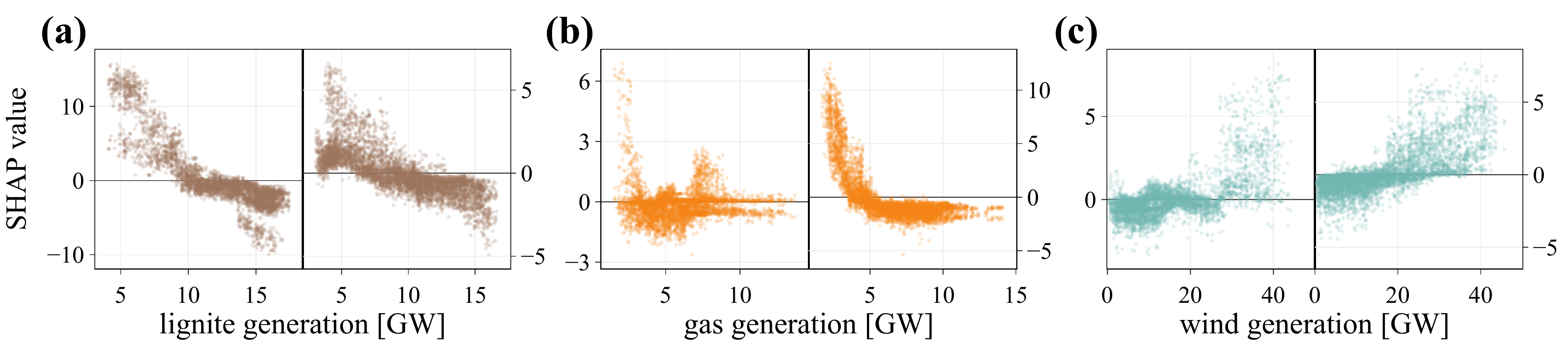}

  \caption{
  Impact of regulatory changes on the SHAP dependency plots of the most important features.
  We display SHAP dependency plots before (left half) and after (right half) the regulatory change on the balancing power market.
  Lignite (a) and gas generation (b) show high positive SHAP values when their share in the generation mix is low. Wind generation (c) shows an opposite dependency.
  The qualitative dependencies of all three types are very similar before and after the regulatory change. 
  }
  \label{shapdependence}
\end{figure*}

\subsection{Models}
To avoid reliance on the peculiarities of individual model classes, we compare the results of two prominent ML algorithms, namely Gradient-Boosted Trees (GBT) and Feedforward Neural Networks (FNN).
We analyze the models in terms of the model-agnostic SHAP values \cite{ lundbergLocalExplanationsGlobal2020}.
For GBTs, the SHAP values can be calculated very efficiently with TreeSHAP~\cite{lundbergConsistentIndividualizedFeature2019, lundbergLocalExplanationsGlobal2020}. Meanwhile, we calculate the SHAP values for FNN with KernelSHAP~\cite{lundbergUnifiedApproachInterpreting2017}. KernelSHAP is an algorithm for approximating Shapley values based on LIME~\cite{ribeiroWhyShouldTrust2016}.
SHAP values are based on the game-theoretic Shapley values~\cite{shapleyValueNPersonGames1953} and inherit their desirable properties (cf.~App.~\ref{app:shap}). They provide both local explanations and global feature importance.

We fit and analyze models separately for the periods before and after the regulatory change. 
For each period, we fit six models to sliding windows, each containing 50\% of the period's data.
This allows us to compare intra- and inter-period variance and reduce the effects of longer-term feature drifts.

The GBT and the FNN have been fitted on the same training and test datasets.
The test datasets each consist of 20\% of four-day blocks that have been randomly selected from the data windows.
For more details about the models we refer to our code~\cite{github}.

\section{Results}
\label{results}
\subsection{Models Change Due to Regulatory Changes}
Both the GBT and the FNN models before and after a regulatory change differ,  see Fig.~\ref{importance_plot}. Inspecting the SHAP importances of the features reveals how certain features become more or less important for the models based on the new regulations. 

The changes observed for the bidding zone split (see Fig.~\ref{importance_plot}a) are straightforward to interpret.
When the price in Austria was equal to the German price, it is evident, that the German residual load is most important for the common price. 
After the division of the bidding zone, Germany, as the neighboring country with the largest population and its heavy industry, still has a strong influence on the electricity price in Austria. However, due to the decoupling of the prices, the influence of the German residual load has decreased.
In the GBT model, Germany's residual load is only 2-3 times more important than the Austrian load, compared to more than 5 times the importance prior to the bidding zone split. In the FNN model, again, the relative importance of Austria greatly increased due to the bidding zone split.

The results for the balancing energy market are more complex to interpret. FNN and GBT models rank the features differently but largely agree on the four most important features of each time period. In addition, for both the GBT and FNN, there is a decrease in the importance of lignite generation and a substantial increase in the importance of gas generation after the end of the mixed price scheme in July 2019.

\subsection{Understanding different models}
SHAP values not only allow us to rank features but also to examine how a particular feature contributes to the prediction of the model depending on its value. 
In the following, we examine particularly important features via their SHAP dependence plots of the GBT models of the balancing market, see Fig.~\ref{shapdependence}.

For the period in which the mixed price system was active, electricity generation by lignite is the most important feature.  
Lignite has substantial market power for negative aFRR (see Fig.~\ref{shapdependence}a).
In general, the SHAP values of the price decrease with lignite generation: If generation is high, lignite power plants can provide negative aFRR at low costs leading to low balancing prices. The dependency plots reveal a change in the market around 10 GW of lignite generation; below 10 GW the dependency is much steeper than above. This is plausible: If a lignite plant is offline or operating at its lower generation threshold, it cannot provide negative aFRR. Hence, lignite plants gradually leave the negative aFRR market as generation falls below 10 GW. Higher bids must be accepted and the price increases.

The dependence of the SHAP values of prices on lignite generation behaves qualitatively similarly even after the end of the mixed price system, but the dependency is generally weaker and the change in slope at 10 GW fades. 
For the gas generation, we observe similar dependencies (see Fig. \ref{shapdependence}b). Here, too, we see an increasing effect on prices when there is little gas generation on the grid. However, this dependency is much more pronounced for gas generation in the second period observed.
For wind generation, we see an opposite dependence in both time periods (see Fig. \ref{shapdependence}c). Here, a high share of wind generation corresponds to high prices on the balancing market. 
A possible explanation: when there is a lot of wind, other, cheaper market participants do not supply sufficient power to the grid to be able to offer negative balancing capacity.

Interestingly, the SHAP dependence plots of all features differ mainly in the magnitude of the SHAP values. The qualitative similarity before and after could indicate that the underlying balancing mechanisms have not changed fundamentally.

\section{Conclusion \& Outlook}
\label{conclusion}
Overall, we have presented two cases of regulatory market changes (bidding zone split and balancing power) and demonstrated how GBT or FNN models change due to changing market rules using SHAP values. For the bidding zone split, we attribute the model changes to the regulatory change with high confidence. However, the balancing market system is a more complex case. Our results indicate that a model change could have been induced by the regulatory change but we cannot exclude the influence of other factors, especially since SHAP values do not establish causal links (cf.~App.~\ref{app:causality}).
 
Concluding, XAI turns modern black-box ML models into an advanced analysis tool for regulatory changes in energy markets. Our method complements human-model-based \cite{tucki2019capacity,pollitt2019european} and common econometric analyses \cite{kochShorttermElectricityTrading2019, eickeElectricityBalancingMarket2021} and can suggest undesirable effects or unintended changes. For example, we revealed the increased importance of lignite during the mixed price system, which is likely undesirable from a regulatory point of view when phasing out coal generation \cite{oei2020coal}.
The two cases presented here serve as a starting point for further research in anomaly detection or explanation and analysis of regulatory changes in the future.

\begin{acknowledgments}
We gratefully acknowledge funding from the Helmholtz Association under grant no. VH-NG-1727 and the Networking Fund through Helmholtz AI. This work was performed as part of the Helmholtz School for Data Science in Life, Earth and Energy (HDS-LEE).
\end{acknowledgments}

\appendix

\section{Data and Code Availability}
Our datasets are obtained from publicly available data sources~\cite{entso-eENTSOETransparencyPlatform, 50hertztransmissiongmbhRegelleistungNet}. The code is accessible on GitHub \cite{github}.
\section{Historic background on the mixed price system}
\label{app:mps}
In the capacity price-based market scheme, the energy prices are completely irrelevant to the awards of the bidding market participants. It is possible to get an award for an offered capacity with extremely high energy prices if the capacity price is low enough. In most cases, this makes activation and thus payment of these prices very unlikely.
But for instance, on October 17, 2017, it happened that a supplier was awarded a large capacity of positive mFRR, which then had to be activated with an energy price of 77,777 €/MWh~\cite{engelhardtEinfuhrungMischpreisverfahrensIm2018}. These prices are usually only three digits.

In its attempt to avoid such events of extreme control prices, the German National Regulator for Energy (Bundesnetzagentur) has introduced the mixed price system. In this system, the mixed price replaces the capacity price in the supplier selection process.

The mixed price is defined as
\begin{equation}
\text{Mixed Price} = \text{Capacity Price} + \alpha \cdot \text{Energy Price},
\end{equation}
where $\alpha$ denotes a quarterly adjusted factor that has been in the single-digit percentage range~\cite{bundesnetzagenturBK618019BeschlussVom2018}.

The mixed price system first came into effect for the delivery date of July 12, 2017. After only two days, the mixed price system was suspended by court order until the auctions on October 16, 2017, and the previous capacity price-based system was temporarily reinstalled~\cite{oberlandesgerichtdusseldorfKart806182018}.

The mixed price system decreased the energy prices as desired, but also entailed undesirable adverse effects. For instance, the low energy prices made it possible to temporarily compensate for forecast errors by depleting the restoration reserves while disregarding available capacities on the intraday market. Thus, the balancing market was misappropriated, and, by unnecessarily activating reserves, the margin for error and the resilience of the system decreased.

In July 2019 the mixed price system was eventually overturned by the Higher Regional Court Düsseldorf and the previous capacity price-based system was reinstated~\cite{oberlandesgerichtdusseldorfKart806182019}. 
\section{SHAP values}
\label{app:shap}

The SHAP values~\cite{lundbergLocalExplanationsGlobal2020} inherit the desirable properties that define the game-theoretic Shapley values~\cite{shapleyValueNPersonGames1953}. 
The \textit{local accuracy} property states that the sum the SHAP values $\phi$ of all features $x_1,\dots,x_n$ matches the models prediction $f(x_1,\dots,x_n)$
\begin{equation}
    f(x_1,\dots,x_n) = \phi_0(f)+\sum_{j=1}^n \phi_j(f,x_1,\dots,x_n)
\end{equation}
where $\phi_0=E[f]$.

We quantify feature importance by averaging the absolute SHAP values per feature and dividing by the sum of all averages
\begin{equation}
    \text{FI}_k = \frac{\langle|\phi_k(f,x_1,\dots,x_n|\rangle_{\text{inputs}}}{\sum_{j=1}^n\langle|\phi_j(f,x_1,\dots,x_n|\rangle_{\text{inputs}}}.
\end{equation}
Accordingly, a feature importance of one would imply that the model relies solely on that particular feature, while a feature importance of zero would imply that the model does not consider the feature at all. For a given model, the feature importances add up to one.
\section{Correlation or causation?}
\label{app:causality}
The causal link between the change in the SHAP value magnitudes and the change in the pricing scheme can of course not be asserted with definitive certainty. At any time, there is a variety of other factors also influencing the energy system. 
For instance, the producer price index (PPI) for lignite increased by about 3 percent from the first to the second period considered, while the PPI for gas decreased by almost 22 percent~\cite{destatisDatenZurEnergiepreisentwicklung2023}.
These price developments have no direct connection to the market changes investigated, but also influence the price at which individual power plant types can offer balancing power.

\bibliography{doc}

\end{document}